\documentclass[aps,prl,twocolumn,amsmath,superscriptaddress,floatfix]{revtex4-2}

\usepackage{graphicx}
\usepackage[dvipsnames]{xcolor} 
\usepackage{epsfig}
\usepackage{epstopdf}

\usepackage{dcolumn}
\usepackage{booktabs}
\usepackage{longtable}
\usepackage{array}

\usepackage{bm}
\usepackage{amssymb}
\usepackage{slashed}

\usepackage{indentfirst}

\graphicspath{{Figures/}}

\usepackage[
    colorlinks,
    citecolor=blue,
    linkcolor=red,
    urlcolor=blue,
    unicode
]{hyperref}

\begin{document}

\title{From Resonance Properties to High-Energy Behavior: The $R$ Ratio from Lattice QCD}

\author{Nan Wang}
\affiliation{State Key Laboratory of Nuclear Physics and Technology, Institute of Quantum Matter, South China Normal University, Guangzhou 510006, China}
\affiliation{Key Laboratory of Atomic and Subatomic Structure and Quantum Control~(MOE), Guangdong-Hong Kong Joint Laboratory of Quantum Matter, Guangzhou 510006, China}
\affiliation{Guangdong Basic Research Center of Excellence for Structure and Fundamental Interactions of Matter, Guangdong Provincial Key Laboratory of Nuclear Science, Guangzhou 510006, China}

\author{Long-cheng Gui}
\affiliation{Department of Physics, Hunan Normal University, Changsha 410081, China}
\affiliation{Synergetic Innovation Center for Quantum Effects and Applications (SICQEA),
Changsha 410081, China}
\affiliation{Key Laboratory of Low-Dimensional Quantum Structures and Quantum Control of Ministry of
Education, Changsha 410081, China}

\author{Jun Hua}
\author{Jian Liang}
\email{jianliang@scnu.edu.cn}
\affiliation{State Key Laboratory of Nuclear Physics and Technology, Institute of Quantum Matter, South China Normal University, Guangzhou 510006, China}
\affiliation{Guangdong Basic Research Center of Excellence for Structure and Fundamental Interactions of Matter, Guangdong Provincial Key Laboratory of Nuclear Science, Guangzhou 510006, China}

\author{Keh-Fei Liu}
\affiliation{Department of Physics and Astronomy, University of Kentucky, Lexington, KY 40506, USA}
\affiliation{Nuclear Science Division, Lawrence Berkeley National Laboratory, Berkeley, CA 94720, USA}

\author{Jun Shi}
\email{jun.shi@scnu.edu.cn}
\affiliation{Guangdong Provincial Key Laboratory of Sensing Physics and System Integration Applications,
School of Physics and Optoelectronic Engineering, Guangdong University of Technology,
Guangzhou, Guangdong 510006, China}

\author{Gen Wang}
\affiliation{Aix-Marseille Universit\'e, Universit\'e de Toulon, CNRS, CPT, IPhU, Marseille, France\\~\\
\includegraphics[scale=0.1]{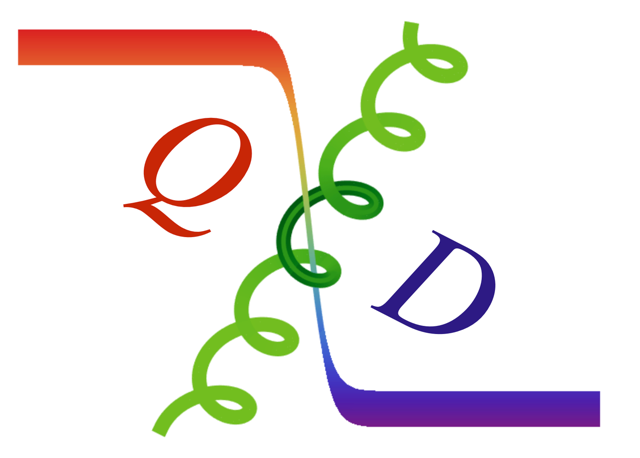}\\
{\rm($\chi$QCD Collaboration)}}

\begin{abstract}
We investigate resonance properties and inclusive spectral behavior through Bayesian reconstruction of the $R$-ratio from Euclidean lattice correlators near the physical pion mass. Without additional Gaussian smearing, the reconstructed spectrum already captures the smooth high-energy behavior observed in experiment. Direct integration of the $\rho$ peak determines $f_\rho$ with approximately $1\%$ statistical precision and close agreement with the experimental reference, providing access to resonance properties without resolving individual finite-volume levels. With a common Gaussian width of $\sigma=0.6$~GeV, the lattice and experimental $R$-ratios show good overall agreement over $0.25$--$2.50$~GeV. Comparisons among ensembles with spatial extents up to approximately $10$~fm demonstrate that finite-volume effects can be controlled across these complementary observables. These results demonstrate a unified framework for studying resonance and inclusive physics from lattice QCD, with potential applications to other hadronic channels, inclusive decay rates, and hadronic tensors.
\end{abstract}

\maketitle

\textit{Introduction}
\label{sec1}
Spectral functions and related inclusive observables encode real-time information, but reconstructing them from finite Euclidean lattice data requires an ill-posed inverse Laplace transform. Numerical approaches include the Maximum Entropy Method~\cite{Asakawa:2000tr}, Backus--Gilbert regularization~\cite{Backus:1968svk}, Bayesian Reconstruction (BR)~\cite{Burnier:2013nla,Rothkopf:2022ctl}, the Hansen--Lupo--Tantalo (HLT) method~\cite{Hansen:2019idp}, machine-learning techniques~\cite{Kades:2019wtd,Buzzicotti:2023qdv}, and other reconstruction methods~\cite{Wang:2021jou,DelDebbio:2024lwm,Bergamaschi:2023xzx}.

The $R$-ratio is defined as the ratio of the cross section for $e^+e^-$ annihilation into hadrons to that into a muon pair. It provides essential input to dispersive determinations of the leading hadronic-vacuum-polarization contribution to the muon anomalous magnetic moment~\cite{Aoyama:2020ynm}. Its inclusive nature involves many excited and multi-hadron states, represented by discrete energy levels in finite volume. Although L\"uscher and Lellouch--L\"uscher methods relate these levels and matrix elements to infinite-volume amplitudes~\cite{Luscher:1985dn,Luscher:1986pf,Luscher:1990ux,Lellouch:2000pv,Lin:2001ek}, a state-by-state analysis becomes increasingly demanding as more coupled channels open.

Spectral reconstruction provides an alternative route for studying inclusive hadronic contributions in finite-volume lattice calculations without resolving individual states. A framework was proposed in Ref.~\cite{Hansen:2017mnd}, in which the infinite-volume spectral function is obtained from its smeared finite-volume counterpart through the ordered double limit
\begin{equation}
\rho(\omega) = \lim_{\sigma\rightarrow0}\lim_{L\rightarrow\infty}\hat{\rho}_L(\omega, \sigma),
\end{equation}
where the infinite-volume limit needs to be taken before the smearing width is sent to zero. At fixed nonzero $\sigma$, the smeared spectral density is a well-defined observable that can be compared directly with experimental data processed using the same resolution kernel. The HLT method constructs smeared observables directly from Euclidean correlators and has been applied by ETMC to calculate the Gaussian-smeared $R$-ratio and compare it with equivalently smeared experimental data~\cite{ExtendedTwistedMassCollaborationETMC:2022sta}. Applications also include inclusive semileptonic $D_s$ decays~\cite{DeSantis:2025qbb,DeSantis:2025yfm}.

In this work, we use the $R$-ratio to demonstrate how spectral reconstruction can access both resonance properties and inclusive spectral behavior with controlled finite-volume effects. For an isolated resonance, finite smearing modifies the line shape, while the $\sigma\to0$ limit is difficult to realize numerically. We therefore study the integrated peak strength, which normalized smearing preserves when leakage outside the integration window is negligible. This allows resonance properties to be determined without resolving individual finite-volume levels or explicitly extrapolating the smearing width to zero. At higher energies, contributions from many states overlap, producing smooth inclusive behavior that can already be studied directly from the reconstructed spectrum. Gaussian smearing provides a complementary comparison with experiment across the full energy range.

We perform the calculation on five ensembles near the physical pion mass, with spatial extents up to approximately $10$~fm. We determine $f_\rho$ with approximately $1\%$ statistical precision and calculate the Gaussian-smeared $R$-ratio up to $2.50$~GeV. The agreement with experiment and demonstrated control of finite-volume effects establish the practical viability of this framework, supporting applications to other hadronic channels, inclusive decay rates, and hadronic tensors.

\textit{Theoretical framework}
\label{sec2}
The zero-momentum Euclidean correlator of the electromagnetic current $J_{\mu}^{\rm em}=\frac{2}{3}\bar{u}\gamma_{\mu}u-\frac{1}{3}\bar{d}\gamma_{\mu}d-\frac{1}{3}\bar{s}\gamma_{\mu}s+\cdots$ can be written as
\begin{equation}
    C(t)=\frac{1}{3}\sum_{i=1}^{3}\int d^3\bm{x}\,
    \left\langle J_i^{\rm em}(t,\bm{x})J_i^{\rm em}(0)\right\rangle.
\end{equation}
Its spectral representation is
\begin{equation}
    C(t) = \int_{0}^{\infty} d\omega\,
    \rho(\omega)e^{-\omega t},
    \label{c2:E}
\end{equation}
where $\rho(\omega)$ is the corresponding spectral function. The optical theorem relates it to the $R$-ratio through
\begin{equation}
    R(\omega) = \frac{12\pi^2}{\omega^2}\rho(\omega),
    \label{R:rho}
\end{equation}
Substituting Eq.~\eqref{R:rho} into Eq.~\eqref{c2:E}, the correlator can equivalently be expressed as
\begin{equation}
    C(t)=\int_{0}^{\infty}d\omega\,
    K(t,\omega)R(\omega),
    \qquad
    K(t,\omega)=\frac{\omega^2e^{-\omega t}}{12\pi^2}.
    \label{c2:reR}
\end{equation}

On a finite-volume lattice, the spectrum is discrete,
\begin{equation}
    \rho_L(\omega)=\sum_n c_n(L)\delta\bigl(\omega-E_n(L)\bigr),
\end{equation}
where $E_n(L)$ and $c_n(L)$ are the finite-volume energy levels and their spectral weights, respectively. Thus, even an exact reconstruction would consist of discrete delta peaks and would not have a well-defined pointwise infinite-volume limit. Smearing defines observables with a well-defined infinite-volume limit at fixed resolution~\cite{Hansen:2017mnd}. For a spectral quantity $F$,
\begin{equation}
    F_{\sigma}(\bar\omega)
    =\int_0^\infty d\omega\,
    \widehat{\mathcal{S}}_{\sigma}(\bar\omega,\omega)F(\omega),
    \label{eq:general_smearing}
\end{equation}
where the kernel is normalized with respect to the output energy,
\begin{equation}
    \int_0^\infty d\bar\omega\,
    \widehat{\mathcal{S}}_{\sigma}(\bar\omega,\omega)=1
    \qquad \text{for every }\omega.
    \label{eq:kernel_normalization}
\end{equation}
With this normalization, the total integral of $F$ is preserved under smearing.

For an isolated resonance, normalized smearing preserves the peak integral if $A$ contains the smeared peak and both leakage from neighboring structures and boundary effects at $\omega=0$ are negligible,
\begin{equation}
    \int_A d\bar\omega\,F_{\sigma}(\bar\omega)
    \simeq \int_A d\omega\,F(\omega).
    \label{eq:peak_area_preservation}
\end{equation}
A derivation is given in the \hyperref[app:area]{End Matter}. Thus, smearing can change the line shape while preserving the integral, allowing finite-volume studies of resonance properties without an explicit $\sigma\to0$ extrapolation.

At higher energies, many contributions overlap. If $F$ varies slowly over the smearing width, finite-width smearing has a much smaller pointwise effect, and $F_{\sigma}$ itself provides a stable inclusive observable at finite resolution.

\textit{Numerical Details}
\label{sec3}
We use overlap valence fermions on five RBC/UKQCD domain-wall-fermion gauge ensembles near the physical pion mass~\cite{RBC:2014ntl,RBC-UKQCD:2008mhs}. Their parameters are summarized in Table~\ref{tab:ensemble}.
With degenerate up and down quarks, the connected electromagnetic correlator is decomposed into light- and strange-quark contributions,
\begin{equation}
C(t) = \frac{5}{9}C^{\mathrm{con}}(t;m_l)+\frac{1}{9}C^{\mathrm{con}}(t;m_s).
\label{eq:flavor_decomposition}
\end{equation}
  Here $m_l=(m_u+m_d)/2$, while $m_s$ is the strange-quark mass. Results at several partially quenched valence-quark masses are interpolated or extrapolated to $m_\pi=135$~MeV, while the physical strange-quark mass is fixed by $m_{\eta_s}=689.89(69)$~MeV. Quark-line disconnected contributions are omitted in the present calculation. We use the renormalized local vector current, with $Z_V$ determined from the Ward identity. Our setup is the same as that in Ref.~\cite{Wang:2022lkq}. Since charm-quark contributions are not included in this calculation, we restrict the displayed energy range to $\omega\leq2.5$~GeV.

\begin{table}[htbp]
\caption{Parameters of the gauge ensembles used in this work. The 24D, 32D, and 48D ensembles are used to study finite-volume effects.}
\label{tab:ensemble}
\begin{tabular}{cccccc}%
\hline  %
\hline  %
    \text{Ensemble} & $L^3\times T$ & $m_\pi^{\mathrm{sea}}~(\mathrm{MeV})$ & $a~(\mathrm{fm})$ & $L~(\mathrm{fm})$ & $N_{\mathrm{cfg}}$ \\
\hline  %
    48I & $48^{3}\times 96$ & 139 & 0.11406(26) & 5.475 & 100 \\
    64I & $64^{3}\times 128$ & 139 & 0.08365(25) & 5.354 & 92 \\
\hline  %
    24D & $24^{3}\times 64$ & 141 & 0.1940(19) & 4.656 & 232 \\
    32D & $32^{3}\times 64$ & 141 & 0.1940(19) & 6.208 & 134 \\
    48D & $48^{3}\times 64$ & 141 & 0.1940(19) & 9.312 & 47 \\
\hline  %
\hline  %
\end{tabular}
\end{table}

We reconstruct the spectral functions using the BR method~\cite{Burnier:2013nla,Rothkopf:2022ctl}, which uses Bayesian prior information to regularize the ill-posed inverse problem and yields a unique maximum-a-posteriori estimate. BR provides comparatively high spectral resolution, although reconstructed structures retain a finite intrinsic width. This broadening is negligible when it is much narrower than the subsequent smearing kernel. Throughout this work, we use a constant default model $m(\omega)=R_0$ to avoid imposing a preferred spectral shape. Because default-model dependence is a principal source of systematic uncertainty in Bayesian spectral reconstruction, we assess it in detail below. We use correlator data from $t/a=1$ and the end time slice varies from $\sim1.5$ to $\sim2.3$ fm. The reconstruction interval $-10\leq\omega\leq25$~GeV is discretized into 4000 points, with the negative-frequency region included for numerical stability. To improve the spectral resolution, the light- and strange-quark contributions are reconstructed separately and then combined using the charge factors in Eq.~\eqref{eq:flavor_decomposition}. The full covariance matrix of the input correlator data is included in the BR likelihood.
Statistical uncertainties are propagated through the analysis using jackknife resampling.
For the smearing applied after the BR reconstruction, we use the normalized Gaussian kernel
\begin{equation}
    \widehat{\mathcal{S}}_{\sigma}(\bar\omega,\omega)
    =\frac{\exp\left[-(\bar\omega-\omega)^2/(2\sigma^2)\right]}
    {\int_0^\infty d\omega'\,
    \exp\left[-(\omega'-\omega)^2/(2\sigma^2)\right]},
\end{equation}
where $\sigma$ is the smearing width.

\begin{figure}[htbp]
    \centering
    \includegraphics[width=0.38\textwidth]{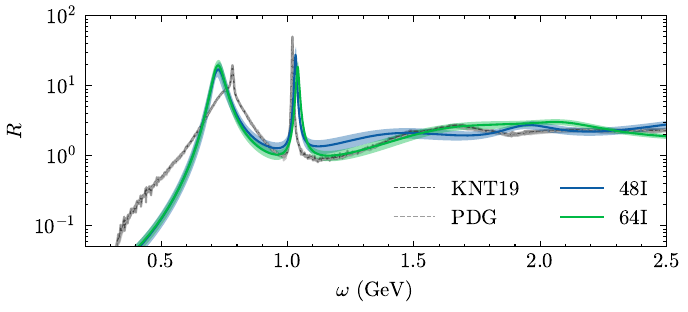}
    \includegraphics[width=0.36\textwidth]{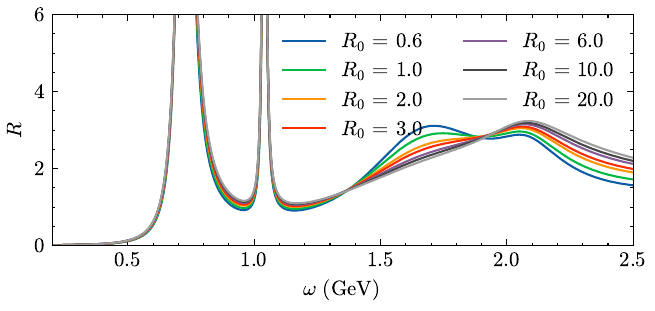}
    \caption{The BR-reconstructed $R$-ratio without Gaussian smearing. The upper panel shows the 48I and 64I results compared with the KNT19~\cite{Keshavarzi:2019abf} and PDG~\cite{ParticleDataGroup:2024cfk} determinations; the lattice bands indicate statistical uncertainties. The lower panel shows a linear-scale view of the 64I result for several constant default models $R_0$.}
    \label{R_not_smear}
\end{figure}

\textit{BR reconstruction without Gaussian smearing}
\label{sec4}
We first examine the BR-reconstructed $R$-ratio without Gaussian smearing. At finite volume, BR's intrinsic resolution broadens the discrete delta-function spectrum into finite-width peaks. As shown in the upper panel of Fig.~\ref{R_not_smear}, the $\rho$ and $\phi$ peaks are clearly resolved at their expected energies. Above approximately $1.4$~GeV, the reconstructed $R$-ratio becomes relatively flat around $R\simeq2$ and agrees qualitatively well with experiment. The lower panel shows the 64I result on a linear scale for several constant default models $R_0$. Below approximately $1.2$~GeV, the reconstructions are nearly independent of $R_0$, indicating that this region is predominantly constrained by the lattice data. Although the pointwise spread among default models grows at higher energies, the broad plateau between 2 and 3 remains relatively stable.

This comparison shows that the high-energy plateau is already present in the BR reconstruction before additional Gaussian smearing is applied. At the finite resolution of the reconstruction, contributions from many excited and multi-hadron states overlap, producing smooth inclusive behavior without the need to resolve individual finite-volume levels.

\begin{figure}[htbp]
    \centering
    \includegraphics[width=0.40\textwidth]{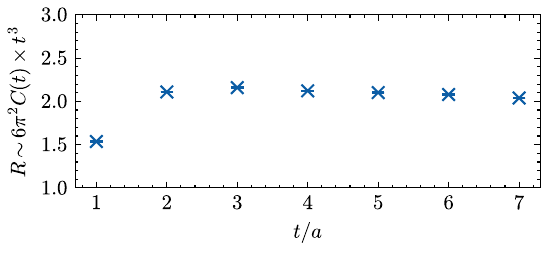}
    \caption{The dimensionless combination $6\pi^2t^3C(t)$ on the 64I ensemble.}
    \label{R_continuous}
\end{figure}

The short-time behavior of the correlator provides a complementary check of this interpretation. For an idealized continuous spectrum with an energy-independent $R$-ratio, $R(\omega)=R_{\rm c}$, Eqs.~\eqref{c2:E} and~\eqref{R:rho} give, for $t>0$,
\begin{equation}
    C^{\rm cont}(t)
    =\frac{R_{\rm c}}{12\pi^2}
    \int_0^\infty d\omega\,\omega^2e^{-\omega t}
    =\frac{R_{\rm c}}{6\pi^2t^3}.
    \label{eq:constant_R_correlator}
\end{equation}
This motivates examining the dimensionless combination $6\pi^2t^3C(t)$. As shown in Fig.~\ref{R_continuous}, this quantity is approximately constant around 2 over $2\lesssim t/a\lesssim7$. Both its magnitude and its time dependence are consistent with the broad $R\simeq2$ plateau observed in the BR reconstruction. The deviation at $t/a=1$ may reflect discretization effects at the shortest time separation. Although the finite-volume spectrum remains discrete, our results suggest that the collective contribution of sufficiently dense levels can be understood through an effective continuous spectrum at short Euclidean times.

\textit{Analysis of the resonance region}
\label{sec5}
We now turn to the resonance region and study the integrated strength of the $\rho$ peak directly from the BR-reconstructed spectra, without additional Gaussian smearing. A conventional Lellouch--L\"uscher analysis requires individual finite-volume energies and matrix elements, together with scattering information to determine the corresponding normalization factors~\cite{Lellouch:2000pv, Lin:2001ek}. These requirements become increasingly demanding as more states and channels contribute. The integrated observables introduced above provide access to the total resonance strength without such a state-by-state analysis. For the $R$-ratio, we apply this strategy to determine $f_\rho$ from the integrated $\rho(770)$ peak.

For the $\rho(770)$, we integrate only the BR-reconstructed light-quark connected electromagnetic contribution, denoted by $R_\ell^{\mathrm{conn}}(\omega)$, over the energy window $[0.5,1.0]$~GeV (the strange-quark contribution in this window is at the per-mille level). We define the isovector integrated strength as
\begin{equation}
    I^R=\frac{9}{10}\int_\rho
    d\omega\,R_\ell^{\mathrm{conn}}(\omega).
    \label{eq:rho_peak_integral}
\end{equation}
The factor $9/10$ removes the $I=0$ contribution from the light-quark connected electromagnetic spectrum.
We express the integrated strength in terms of $f_\rho$ using
\begin{equation}
    f_\rho=\left(\frac{m_\rho I^R}{6\pi^2}\right)^{1/2}.
    \label{eq:rho_decay_constant_integral}
\end{equation}
This expression gives the decay constant in the narrow-width limit. We obtain the experimental counterpart by integrating over the same window, with the $\omega$ contribution removed using the parameterization described in the \hyperref[app:rho_reference]{End Matter}.

\begin{figure}[htbp]
    \centering    
    \includegraphics[width=0.45\textwidth]{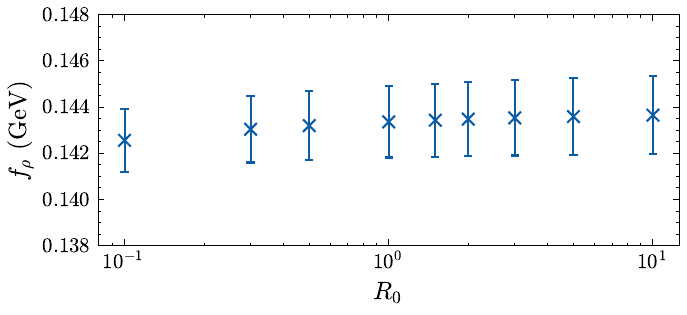}
    \caption{Default-model dependence of $f_\rho$ on the 64I ensemble.}
    \label{rho_f_prior}
\end{figure}

For the lattice determination of $f_\rho$, we first examine the default-model dependence. We vary the constant default model $R_0$ over $0.1\leq R_0\leq10$ on the 64I ensemble. As shown in Fig.~\ref{rho_f_prior}, $f_\rho$ depends only weakly on $R_0$, with its central value becoming less sensitive to $R_0$ at larger values.
We take the result at $R_0=2$ as the central value, motivated by the quark-model prediction $R=2$ for three light flavors. The observed default-model variation is smaller than the statistical uncertainty, so we do not assign a separate systematic uncertainty from this source.

\begin{figure}[htbp]
    \centering    
    \includegraphics[width=0.45\textwidth]{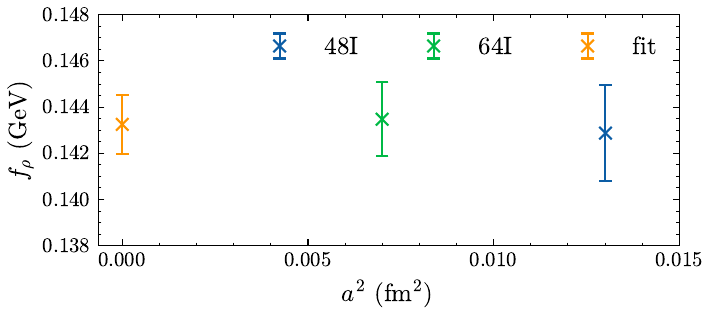}
    \par\smallskip
    \includegraphics[width=0.45\textwidth]{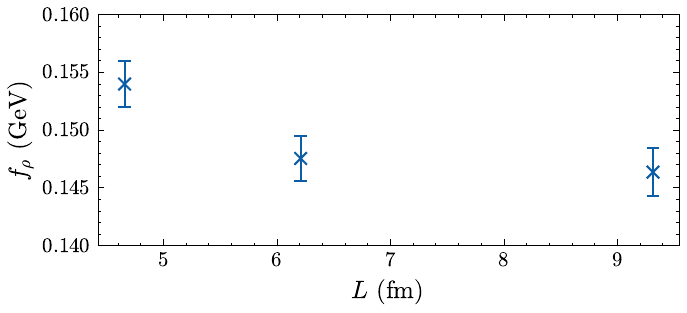}
    \caption{Lattice-spacing (upper panel) and finite-volume (lower panel) dependence of $f_\rho$.}
    \label{rho_f_volume_spacing}
\end{figure}

We then assess lattice-spacing and finite-volume effects through the comparisons shown in Fig.~\ref{rho_f_volume_spacing}. In the upper panel, the 48I and 64I results, obtained at similar spatial extents, agree within statistical uncertainties and show no resolved lattice-spacing dependence. A constant fit gives $f_\rho=0.1433(13)$~GeV, which we take as the central value. Its difference from the finer-lattice result is much smaller than the statistical uncertainty, so we do not assign a separate discretization uncertainty. The lower panel compares the 24D, 32D, and 48D results at a common lattice spacing and different spatial volumes. The 32D and 48D results are consistent within statistical uncertainties, indicating no resolved volume dependence for $L\gtrsim6$~fm at the current precision. This demonstrates that resonance properties can be studied through the integrated spectral strength with controlled finite-volume effects. Since the spatial extents of 48I and 64I lie between those of 24D and 32D, we conservatively assign a $4\%$ finite-volume systematic uncertainty to their results, based on the relative difference between the 24D and 32D determinations.

Including the finite-volume systematic uncertainty, we quote
\begin{equation}
    f_\rho=0.1433(13)(57)~\mathrm{GeV},
\end{equation}
where the first uncertainty is statistical and the second is the finite-volume systematic uncertainty.
The corresponding experimental reference, based on the KNT19 data~\cite{Keshavarzi:2019abf}, is $f_\rho^{\mathrm{exp}}=0.14145$~GeV. The lattice central value is approximately $1.3\%$ above this reference and is consistent with it within the quoted lattice uncertainties. This close agreement with experiment further supports the viability of our approach and highlights its value for quantitative studies of resonance properties in lattice QCD.

\begin{figure}[htbp]
    \centering  
    \includegraphics[width=0.45\textwidth]{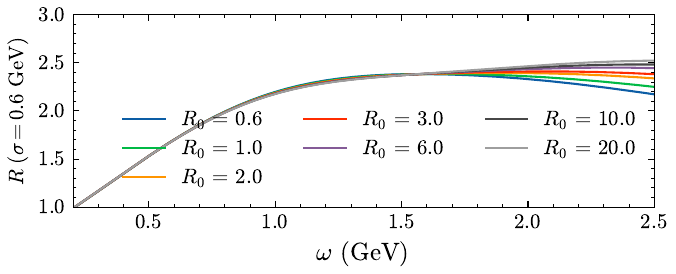}
    \caption{Default-model dependence of $R_\sigma(\omega)$ on the 64I ensemble for $\sigma=0.6$~GeV.}
    \label{R_64I_smear}
\end{figure}

\textit{Gaussian-smeared $R$-ratio}
\label{sec6}
We now turn to the Gaussian-smeared $R$-ratio, applying the same smearing kernel to the lattice and experimental spectra for a direct comparison. We first examine the default-model dependence on the 64I ensemble, taking $\sigma=0.6$~GeV as an example. Figure~\ref{R_64I_smear} shows results for $0.6\leq R_0\leq20$. Below approximately $1.8$~GeV, the results show no appreciable default-model dependence, indicating that the reconstruction is strongly constrained by the lattice data. At higher energies, the default-model dependence becomes visible as the data constraints weaken, but a meaningful determination remains possible up to $2.5$~GeV when this uncertainty is taken into account. For all smearing widths, we use $R_0=2$ for the central results and take half the absolute difference between the $R_0=0.6$ and $R_0=6$ results as the default-model systematic uncertainty.

\begin{figure}[htbp]
    \centering    
    \includegraphics[width=0.45\textwidth]{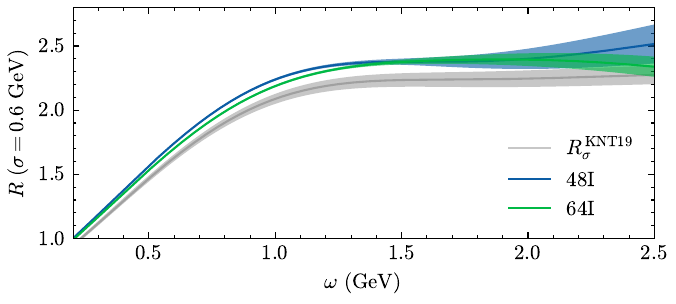}
    \includegraphics[width=0.45\textwidth]{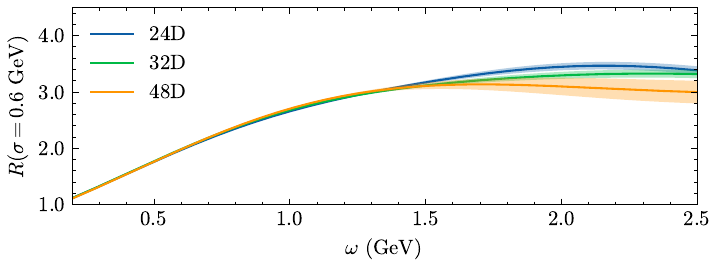}
    \caption{Lattice-spacing (upper panel) and finite-volume (lower panel) dependence of $R_\sigma(\omega)$ for $\sigma=0.6$~GeV. Bands indicate statistical uncertainties.}
    \label{R_sm_dot6_finite_volume}
\end{figure}

\begin{figure}[htbp]
    \centering    
    \includegraphics[width=0.45\textwidth]{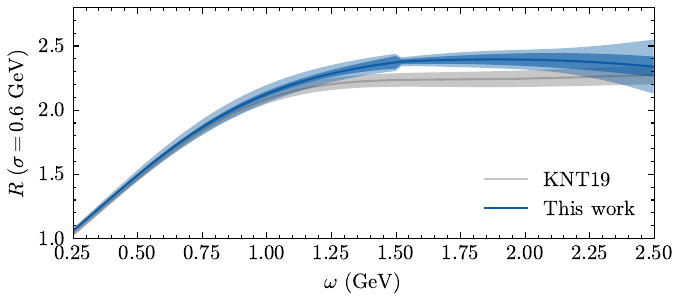}
    \caption{$R_\sigma(\omega)$ at $\sigma=0.6$~GeV. Dark and light blue bands indicate the lattice statistical and total uncertainties, respectively; the gray band shows the KNT19 result.}
    \label{R_final}
\end{figure}

We assess lattice-spacing effects using the 48I and 64I results shown in the upper panel of Fig.~\ref{R_sm_dot6_finite_volume}. Below $1.52$~GeV, we obtain the central value and statistical uncertainty from a linear extrapolation in $a^2$. At higher energies, where the lattice-spacing dependence is not statistically resolved, we use the 64I result and its statistical uncertainty. Throughout the energy range, we conservatively assign the absolute difference between the 48I and 64I results as the discretization systematic uncertainty. The switch between the two prescriptions accounts for the apparent discontinuity in the uncertainty band near $1.52$~GeV in Fig.~\ref{R_final}.
We also examine finite-volume effects using the 24D, 32D, and 48D ensembles, shown in the lower panel of Fig.~\ref{R_sm_dot6_finite_volume}. At low energies, no volume dependence is resolved despite the small statistical uncertainties. At higher energies, the 24D and 32D results remain consistent within statistical uncertainties. We therefore do not assign a separate finite-volume systematic uncertainty at the current precision.

Figure~\ref{R_final} shows our final results for $\sigma=0.6$~GeV over $0.25\leq\omega\leq2.50$~GeV, together with the KNT19 data~\cite{Keshavarzi:2019abf} smeared using the same kernel. The total lattice uncertainty is obtained by adding the statistical, default-model, and discretization uncertainties in quadrature. The lattice results agree closely with experiment below $1.2$~GeV with small total uncertainties, and are also consistent with experiment above $2$~GeV. In the intermediate region, the lattice values are higher, probably reflecting the omission of quark-line disconnected contributions.

Results for $\sigma=0.5$ and $0.7$~GeV are presented in the \hyperref[app:smearing_widths]{End Matter} and show a similar pattern of agreement with experiment.

\textit{Conclusions}
\label{sec7}
In this work, we have calculated the $R$-ratio from lattice QCD using BR spectral reconstruction to investigate both resonance properties and inclusive spectral behavior. We find that, without additional Gaussian smearing, the reconstructed $R$-ratio already captures the smooth high-energy behavior observed in experiment. In the resonance region, direct integration of the reconstructed $\rho$ peak determines $f_\rho$ with approximately $1\%$ statistical precision and close agreement with the experimental reference. This demonstrates that resonance properties can be studied through integrated spectra without resolving individual finite-volume levels. To compare the spectra across the full energy range, we then apply Gaussian smearing and obtain good overall agreement with experiment up to $2.5$~GeV.

We emphasize that finite-volume effects are well controlled in all these calculations. Together, these results demonstrate a unified framework for studying resonance and inclusive physics from Euclidean lattice correlators. Extending this framework to other hadronic channels, inclusive decay rates, and hadronic tensors would broaden the scope of first-principles studies of resonance properties and complex hadronic dynamics without resolving individual finite-volume levels.

\textit{Acknowledgments}
\label{sec8}
We thank the RBC/UKQCD collaboration for sharing their DWF gauge configurations. The numerical calculations are performed on the Southern Nuclear Science Computing Center (SNSC) of South China Normal University.
This work is supported by the National Natural Science Foundation of China (NSFC) under Grants No.\ 12575085, No.\ 12222503, No.\ 12635009 and No.\ 12675103.

\bibliographystyle{apsrev4-2}
\bibliography{lib.bib}

\clearpage
\begin{center}
    \textbf{End Matter}
\end{center}
\label{em:start}
\appendix

\newcommand{\prlendmatterheading}[2]{%
    \par\medskip\refstepcounter{section}\label{#1}%
    \noindent\textit{#2} \ignorespaces}

\prlendmatterheading{app:area}{Area preservation for an isolated peak}
The normalization condition in Eq.~\eqref{eq:kernel_normalization} first guarantees global area preservation. Exchanging the order of integration gives
\begin{align}
    \int_0^\infty d\bar\omega\,F_{\sigma}(\bar\omega)
    &=\int_0^\infty d\omega\,F(\omega)
    \int_0^\infty d\bar\omega\,
    \widehat{\mathcal{S}}_{\sigma}(\bar\omega,\omega) \notag\\
    &=\int_0^\infty d\omega\,F(\omega).
    \label{eq:global_area_preservation}
\end{align}

For the local version of this result, let $F_{\rm pk}(\omega)$ denote an isolated peak with support contained in an interval $A_0$, and choose a wider integration window $A$ that contains its smeared profile. The smeared contribution from this peak is
\begin{equation}
    F_{{\rm pk},\sigma}(\bar\omega)
    =\int_{A_0}d\omega\,
    \widehat{\mathcal{S}}_{\sigma}(\bar\omega,\omega)
    F_{\rm pk}(\omega).
\end{equation}
Its integral over $A$ can be written as
\begin{align}
    \int_A d\bar\omega\,F_{{\rm pk},\sigma}(\bar\omega)
    =\int_{A_0}d\omega\,F_{\rm pk}(\omega)
    \left[\int_A d\bar\omega\,
    \widehat{\mathcal{S}}_{\sigma}(\bar\omega,\omega)\right].
\end{align}
If the kernel tails outside $A$ and contamination from neighboring structures are negligible, the expression in brackets is approximately unity. Since the original peak is fully contained in $A$, this yields Eq.~\eqref{eq:peak_area_preservation}. Thus, normalized smearing redistributes an isolated peak without changing its integral. Normalizing the kernel over the physical output-energy domain, as in Eq.~\eqref{eq:kernel_normalization}, also accounts for the boundary at $\omega=0$; the local result requires the chosen window to contain the smeared peak with negligible leakage.

\begin{figure}[htbp]
    \centering    
    \includegraphics[width=0.40\textwidth]{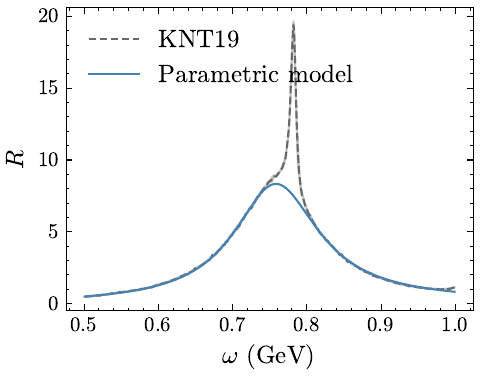}
    \caption{The $R$-ratio in the $\rho$ region. The gray dashed curve and band show the KNT19 data, including the $\omega$ peak; the blue solid curve shows the parameterized $\rho$ contribution.}
    \label{R_rho_model}
\end{figure}

\prlendmatterheading{app:rho_reference}{Experimental reference for the $\rho$ resonance}
The KNT19 data retain the narrow $\omega$ peak, as shown in Fig.~\ref{R_rho_model}. We remove this contribution from the experimental reference through a parameterization of the broad $\rho$ contribution. The factor $9/10$ in Eq.~\eqref{eq:rho_peak_integral} applies only to the lattice light-quark connected spectrum and is not used for the experimental data.

To describe the smooth peak in the $\rho$ region, we parameterize the pion electromagnetic form factor as
\begin{equation}
    |F_\pi(\omega)|^2
    =v_0+\frac{v_1}{(\omega/m_\pi-v_2)^2+v_3^2}.
    \label{eq:rho_parametric_form_factor}
\end{equation}
Here $v_0$ describes a smooth background, while $v_1$, $v_2$, and $v_3$ control the peak strength, location, and width. The corresponding two-pion contribution is~\cite{Jegerlehner:2009ry,Meyer:2011um}
\begin{equation}
    R_{\pi\pi}(\omega)
    =\frac{1}{4}\left(1-\frac{4m_\pi^2}{\omega^2}\right)^{3/2}
    |F_\pi(\omega)|^2.
    \label{R_sp_rho}
\end{equation}

The fit uses the KNT19 data in $[0.5,0.74]$ and $[0.82,1.0]$~GeV, excluding the $\omega$-peak region. The fitted smooth parameterization is then integrated over the full $[0.5,1.0]$~GeV window and converted to $f_\rho^{\mathrm{exp}}$ using Eq.~\eqref{eq:rho_decay_constant_integral}. The same reference mass is used for the lattice and experimental determinations.

\prlendmatterheading{app:smearing_widths}{Additional smearing widths}
The analysis and uncertainty prescriptions are the same as those used for $\sigma=0.6$~GeV in the main text.

\begin{figure}[htbp]
    \centering
    \includegraphics[width=0.45\textwidth]{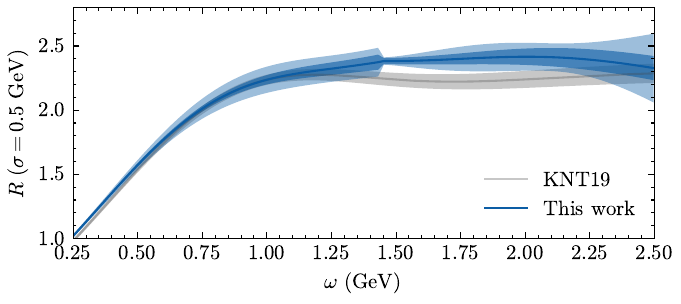}
    \par\smallskip
    \includegraphics[width=0.45\textwidth]{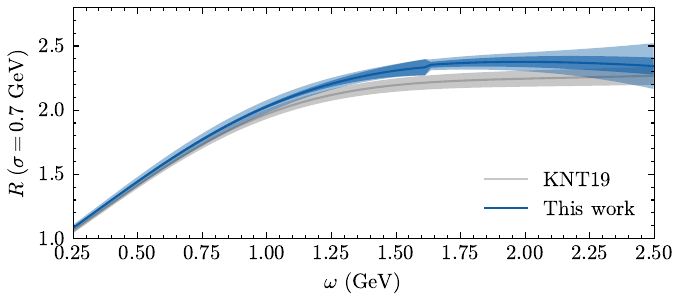}
    \caption{$R_\sigma(\omega)$ at $\sigma=0.5$~GeV (upper panel) and $0.7$~GeV (lower panel). Dark and light blue bands indicate the lattice statistical and total uncertainties, respectively; gray bands show the KNT19 results.}
    \label{R_final_other_widths}
\end{figure}

\label{em:end}
\end{document}